**Research Article**

**Establishment of a diagnostic model to distinguish coronavirus disease 2019 from influenza A based on laboratory findings**

**Running title:** A diagnostic model to distinguish coronavirus disease 2019 from influenza A

Dongyang Xing[1], Suyan Tian[2], Yukun Chen[3], Jinmei Wang[4], Xuejuan Sun[5], Shanji Li[6], Jiancheng Xu[2]

[1] Department of Laboratory Medicine, First Hospital of Jilin University, Changchun 130021, China

[2] Division of Clinical Research, First Hospital of Jilin University, Changchun 130021, China

  Department of Laboratory Medicine, First Hospital of Jilin University, Changchun 130021, China

[3] Department of Infection Control, First Hospital of Jilin University, Changchun 130021, China

[4] Department of Laboratory Medicine, Siping Infectious Disease Hospital, Siping 136099, China

[5] Department of Laboratory Medicine, Changchun Infectious Disease Hospital, Changchun 130123, China

[6] Department of Laboratory Medicine, Jilin Infectious Disease Hospital, Jilin , 132011 China

**Correspondence author:**

Jiancheng Xu, MD, PhD;

Department of Laboratory Medicine, The First Bethune Hospital of Jilin University；1 Xinmin

Street，Changchun, 130021

Phone: +86-431-8878-2595

Fax: +86-431-8878-6169

E-mail: xjc@jlu.edu.cn

**Abstract**

**Background:** Coronavirus disease 2019 (COVID-19) and Influenza A are common disease caused by viral infection. The clinical symptoms and transmission routes of the two diseases are similar. However, there are no relevant studies on laboratory diagnostic models to discriminate COVID-19 and influenza A. This study aims at establishing a signature of laboratory findings to tell patients with COVID-19 apart from those with influenza A perfectly.

**Materials:** In this study, 56 COVID-19 patients and 54 influenza A patients were included. Laboratory findings, epidemiological characteristics and demographic data were obtained from electronic medical record databases. Elastic network models, followed by a stepwise logistic regression model were implemented to identify indicators capable of discriminating COVID-19 and influenza A. A nomogram is diagramed to show the resulting discriminative model.

**Results:** The majority of hematological and biochemical parameters in COVID-19 patients were significantly different from those in influenza A patients. In the final model, albumin/globulin (A/G), total bilirubin (TBIL) and erythrocyte specific volume (HCT) were selected as predictors. Using an external dataset, the model was validated to perform well.

**Conclusion:** A diagnostic model of laboratory findings was established, in which A/G, TBIL and HCT were included as highly relevant indicators for the segmentation of COVID-19 and influenza A, providing a complimentary means for the precise diagnosis of these two diseases.

Page 4

**Introduction**

Since COVID-19 emerged in 2019, the disease has spread rapidly around the world and attracted global attention. The epidemic of severe acute respiratory syndrome coronavirus 2 (SARS-CoV-2) was declared an international public health emergency by the World Health Organization (WHO). The early typical clinical symptoms of the disease are fever, respiratory symptoms and muscle pain (Mark R. Geier and Geier 2020). In severe cases, acute respiratory distress syndrome (ARDS) or even respiratory failure (RF) can develop.

During winter, viral infectious diseases gradually enter a high incidence period. As a common seasonal influenza, influenza A tends to be prevalent in winter (Lei et al. 2018). This virus is transmitted through three main routes (Sullivan et al. 2010) (contact transmission, droplet transmission and airborne transmission), which can lead to a wide range of human-to-human transmission. The early clinical symptoms of influenza A patients include fever, headache, muscle pain and dyspnea (Schoen et al. 2019), which are similar to the early onset of COVID-19. Hashemi et al. found the co-infection of SARS-CoV-2 with other respiratory viruses(Hashemi et al. 2020). Dr. Nanshan Zhong (China Health Science and Technology Innovation and Development Conference reports. http://www.sz.gov.cn/cn/xxgk/zfxxgj/zwdt/content/post_8307353.html Accessed November 27 2020.) has also reported that four patients were infected with influenza A and COVID-19 simultaneously, and thus, an accurate and efficient segmentation of influenza A and COVID-19 is of crucial importance.

Some studies have reported that the count of white blood cell (WBC), lymphocyte (LY) and platelet (PLT) were decreased, and the levels of alanine aminotransferase (ALT) and aspartate aminotransferase (AST) were increased in H1N1 and H5N1 patients (Gao et al. 2013; Natalie L. Cobb et al. 2020). COVID-19 patients can also exhibit these changes (Sun et al. 2020; Tian et al. 2020). The important laboratory findings have shown similar trends in COVID-19 and influenza A. In addition, the clinical symptoms, transmission routes and onset seasons are similar, which suggests that it is necessary to distinguish patients infected with SARS-CoV-2 and those with influenza A virus. In this paper, we aimed to establish a diagnostic model of laboratory findings to distinguish patients with COVID-19 and those with influenza A, and to identify highly relevant indicators associated with the phenotype of interest. The establishment of the model will not only help in the timely diagnosis of COVID-19 and influenza A patients but also reduce the occurrence of complications and contribute to clinical treatment.

**Materials and Methods**

**Patient Selection and Data Sources**

In the training set, the patients were recruited from three designated tertiary hospitals in Jilin Province, including 56 COVID-19 patients recruited from the First Hospital of Jilin University ($n = 3$), Changchun Infectious Disease Hospital ($n = 39$) and Siping Infectious Disease Hospital ($n = 14$) from January to March 2020, and 54 patients with influenza A recruited from the First Hospital of Jilin University from December 2019 to March 2020. The 24 COVID-19 patients recruited from Jilin Infectious Disease Hospital from January to May 2020, and 30 influenza A patients recruited from The First Hospital of Jilin University from December 2018 to April 2019 were combined together as one for the purpose of external validation.

Laboratory findings, clinical symptoms and demographic data were retrieved from electronic medical records. The data were extracted from local databases by experienced medical professionals, and two researchers examined the data independently. Specimens were collected according to the standards of each laboratory. Nasal swabs, pharyngeal swabs and venous blood samples were generally collected. The exclusion criteria for patients in this study were as follows: 1. infection with other bacteria or common viruses; 2. organic diseases such as heart, lung, liver, gallbladder, kidney, and blood diseases; and 3. pregnancy or lactation.

**Laboratory Tests**

Laboratory confirmation of the COVID-19 patients was completed by Changchun Centers for Disease Control and Prevention (CDC), Siping CDC, and Jilin CDC. The specimens of the patients suspected to be positive for COVID-19 were transported to the Jilin Provincial CDC for confirmation. Laboratory confirmation of influenza A patients was completed by the Changchun CDC, and the main types of influenza A patients were those infected with A (H1N1) pdm 09 and H3N2. SARS-CoV-2 was detected by reverse transcription polymerase chain reaction (RT-PCR) (Shanghai Biogerm Medical Technology Co., Ltd., Shanghai GeneoDx Biotech Co., Ltd.). Influenza A virus was also detected by RT-PCR (Shanghai GeneoDx Biotech Co., Ltd). Hematological and biochemical tests were performed on specimens from COVID-19 and influenza A patients in the four laboratories. Supplementary Table X1 presents the relevant information on the testing equipment in each hospital. Four hospitals participated in and passed the external quality assessment and proficiency testing of the clinical laboratory center of Jilin Province, and the testing kits and equipment of the four laboratories were matched. All the doctors, technicians and nurses in this study received unified training from the Jilin Provincial Health Commission. In this paper, laboratory findings for COVID-19 and influenza A patients were compared with those of health industry standards in the People's Republic of China, including common biochemical and hematological parameter reference intervals (RIs) used in clinical settings.

**Statistical Analysis**

**Identification of diagnostic indices**

To identify the indices that have diagnostic value for distinguishing between COVID-19 and influenza A, we carried out the following machine learning procedure. First, the laboratory findings during the first five days of the hospital stay were averaged and then used as potential attributes for feature selection. Elastic net models were used for the first round of selection, in which the tuning parameter alpha was set at 0.6 and the optimal cutoff of lambda was chosen by performing 10-fold cross-validations. With different random seeds, the elastic net models were fit for 200 times. Then, the frequencies of the indices selected by these 200 models were calculated. When the indices with high frequencies were selected ($> 90\%$), the pairwise correlation coefficients for each pair were calculated. To eliminate redundant indices and to avoid collinearity issues among the highly correlated indices, those with the least biological meaning were excluded. For the second-round feature selection modeling, stepwise logistic regression models with AIC as the selection criterions were utilized. Finally, a ridge logistic regression model was fit with the selected indices as predictors. A nomogram was diagramed to graphically elucidate the final model. Using an independent dataset as an external validation set, the area under the receiver operating characteristic (ROC) curve (AUC) metric was calculated to evaluate the predictive performance of the final model.

**Comparison at baseline**

Chi-square test is applied to the processing of basic information data, and non-parametric method is applied to calculate the reference value of 2.5% -95% for the RIs of laboratory test index. The statistical processing by using IBM SPSS Statistics 26 software.

**Results**

**Demographics**

In this study, 56 COVID-19 patients (aged 10 to 87 years) and 54 influenza A patients (aged 23 to 87 years) were included. The demographic characteristics of these two groups are presented in Table 1. Hospitalization time for COVID-19 patients was longer. There was no significant difference between two groups in males/females ratio, but the median age of COVID-19 patients was lower than that of influenza A patients. Fever, cough and fatigue were the common clinical symptoms of COVID-19 patients. The common clinical symptoms of influenza A patients were fever, cough and dyspnea.

**Laboratory Findings at Hospital Admission**

In this study, laboratory findings were performed on the hematological and biochemical parameters of 56 COVID-19 patients and 54 influenza A patients. The first laboratory findings from the hospital patients are shown in Table 2. Among the leukocyte parameters, the count of WBC, neutrophil (NE), and monocyte (MO) in the influenza A patients were significantly higher than those in the COVID-19 patients ($P < 0.001$); however, there was no significant difference in lymphocyte (LY), eosinophilic granulocyte (EO) and basophilic granulocyte (BA) ($P > 0.05$). In terms of erythrocyte parameters, the count of RBC, hemoglobin (HGB), HCT and mean erythrocyte hemoglobin concentration (MCHC) in the COVID-19 patients were all higher than those in the influenza A patients, while the RBC distribution width (RDW) was lower than that of influenza A patients ($P < 0.001$). Regarding PLT parameters, the mean platelet volume (MPV) of the COVID-19 patients was significantly reduced ($P < 0.001$). Regarding the biochemical parameters, there was a significant difference in the levels of other parameters except for aspartate aminotransferase (ALT), creatinine (Cr) and glucose (GLu). Among the electrolyte parameters, potassium (K+) and sodium (Na+) levels in the COVID-19 patients were higher than those in the influenza A patients, while the chlorine (Cl$^-$) levels in the COVID-19 patients were lower.

**Diagnostic Indices to Discriminate between COVID-19 and Influenza A**

Following the feature selection procedure given in the Materials and Methods section, the diagnostic signature to distinguish COVID-19 and influenza A was constructed. In the first round of feature selection, EO, MO, RBC, HGB, HCT, MPV, GGT, TP, ALB, GLB, A/G, TBIL, indirect bilirubin (IBIL), urea nitrogen (UR), $CO_2$-CP, Cl$^-$, and age had frequencies >

90% over 200 elastic net models (the plot to select the optimal tuning parameter lambda for a single run of elastic net is shown in Figure 1A). Then, their correlation plot was diagrammed (Figure 1B) to examine the potential redundant indices. Next, the clinical significance of each index was thoroughly examined by an expert specialist (Dr. Xu) to determine which highly correlated indices should be included. Then, the second-round feature selection was carried out, yielding A/G, TBIL and HCT as the resulting best subset model. The final model is represented by,

$$\text{logit}(p_i) = -15.28 + 7.21 * A/G_i - 0.08 * TBIL_i + 17.03 * HCT_i$$

where $p_i$ represents the probability of being infected with COVID-19 for patient *i*, with the corresponding p-values for A/G, TBIL and HCT of <0.001, 0.014 and 0.037, respectively. A graphical illustration of this model with the aid of a nomograph is presented in Figure 3. This model has been demonstrated to have a satisfactory predictive performance to discriminate between COVID-19 and influenza A (AUC=0.844) using an external validation set, as shown in Figure 4.

**Discussion**

Until now, few studies have been carried out to compare the laboratory findings and clinical symptoms of COVID-19 and influenza A patients. For instance, Li (Li et al. 2020) and Natalie(Cobb et al. 2020) et al. found significant differences in the counts of WBC and NE but no significant differences in AST, ALT, Cr or other indicators. Tang et al. (Xiao Tang et al. 2020) showed no difference in the counts of WBC, NE and LY, but there were significant differences in PLT, ALB and AST.

Significant differences in most laboratory findings between COVID-19 and influenza A patients at the onset of diseases were found. For example, the count of WBC and NE decreased significantly in COVID-19 patients, which is consistent with the findings of Chen's study (Chen et al. 2020) and Guan's study (Guan et al. 2020). On the other hand, several studies reported that WBC and NE were increased on the first day of infection in influenza A patients such as (Ma S et al. 2020; Wang et al. 2014). LY was decreased more significantly in influenza A patients, compared to COVID-19 patients. Chen (Chen et al. 2013)et al. found that patients infected with H7N9 had lymphopenia. Another study showed (Ong et al. 2009) that LY had a high specificity in the laboratory diagnosis of influenza A and could improve the detection rate of H1N1 patients relatively. Flick (Flick et al. 2014)et al. also proposed that fever (body temperature > 38℃) and changes in leukocyte parameters in influenza A patients are diagnostic criteria that can increase the sensitivity of clinical diagnosis to 86.4%. However, several studies (Kumar et al. 2009; Nseir et al. 2008; Perez-Padilla et al. 2009; Rodriguez-Noriega et al. 2010)have suggested that leukocyte parameters may be more helpful in identifying viral or bacterial respiratory infections. In addition, the erythrocyte parameters of patients in both groups were also changed. Specifically, the RBC, HGB and HCT counts of the influenza A patients decreased more significantly than those of the COVID-19 patients, while RDW was higher than that of the COVID-19 patients. Salvagno (Salvagno et al. 2015) et al. reported that RDW was an important indicator of red blood cell homeostasis and impaired red blood cell production. In addition, elevated RDW was also a marker of inflammation and oxidation state. In the early stage of infection, H1N1 patients had a high frequency of fever and pneumonia(Weiss and Goodnough 2005), and thus the RDW of the influenza A patients was significantly increased. In addition, the count of PLT in influenza A and COVID-19 patients were significantly reduced. Abelleira et al. found in control study of patients with influenza A that the count of PLT in the case group was lower than that in the control group (Abelleira et al. 2019). Chen (Chen et al. 2020)et al. and Guan (Guan et al. 2020) et al. both confirmed that PLT level was reduced in COVID-19 patients.

There were also significant changes in biochemical markers. Some studies have reported that patients with COVID-19(Chen et al. 2020; Cheng et al. 2020; Skevaki et al. 2020) and influenza A (To et al. 2014; Zhang et al. 2014) have different degrees of kidney injury and liver injury for unknown reasons. The AST, ALT and GGT levels of most patients in the two groups were all higher than the upper limit of the RIs, while the TP, ALB and GLB levels showed a decreasing trend. These results were consistent with the studies of Romina (Abelleira et al. 2019; Gao et al. 2013)et al. Carbon dioxide combining power  (CO2-CP) represents the level of bicarbonate in plasma. The average measured value in the COVID-19 patients was lower than the lower limit of the RIs. Metabolic acidosis was reported in patients infected with SARS-CoV-2(Xu Cheng et al. 2020) with severe disease, which directly reduced the CO2-CP. Other studies have also confirmed(Gao et al. 2013; Zhang et al. 2014) that patients with H1N1 had clinical symptoms of hypoxemia and reduced partial pressure of carbon dioxide, which were associated with reduced CO2-CP. At the same time, the $K^+$, $Na^+$ and Cl+ levels of patients in the two groups were also decreased. Gao (Gao et al. 2013) and Chen (Chen et al. 2020) et al. mentioned that both COVID-19 and influenza A patients suffered vomiting and diarrhea, and more than one-half of influenza A patients were reported to have hypokalemia and hyponatremia (Zhang et al. 2014). The specific reasons are not clear, but it is currently believed that electrolyte parameter changes may be related to the above clinical symptoms.

Although some laboratory findings were found to be different in the two diseases, their variation trends were similar. Therefore, to better classify the two diseases, this study aimed to establish a diagnostic model according to laboratory findings for COVID-19 and influenza A and then select the most representative indicators for the clinical identification of the two viral infections. Using machine learning methods, we showed that the three laboratory findings of A/G, TBIL and HCT possess predictive capacity to discriminate the two diseases ($P$<0.001, 0.014 and 0.037, respectively). Studies have found that influenza A patients exhibit hypoproteinemia and hypoalbuminemia (Zhang et al. 2014). Tang (Xiao Tang et al. 2020)et al. reported that the ALB level of influenza A patients was significantly lower than that of COVID-19 patients. In this study, the GLB level of influenza A patients was higher than that of COVID-19 patients. These conclusions indirectly proved that the A/G ratio in influenza A patients was decreased significantly, which was consistent with the results of this study. In addition, TBIL level in influenza A patients were significantly higher than those in COVID-19 patients. A cohort study by Zhang(YiMin Zhang et al. 2016) and Tang(Xiao Tang et al. 2020) et al. confirmed that the TBIL level in influenza A patients was increased significantly. This may also be due to liver injury caused by clinical drugs, which somehow influence TBIL level. The drug oseltamivir is commonly used for the treatment of influenza A virus, which is metabolized in vitro by liver esterase. The frequent use of oseltamivir reduces the level of liver esterase and leads to drug residues in the body of patients, thereby causing liver damage(Shengbo Fang et al. 2018). Hematocrit (HCT) refers to the volume ratio of sinking red blood cells to whole blood measured after centrifugal precipitation of a certain amount of whole blood treated with anticoagulant, which indirectly reflects the number and volume of red blood cells. In this study, the HCT level in the influenza A patients was lower than that in the COVID-19 patients, and the RBC count was significantly lower than that in the COVID-19 patients. It was recently confirmed that influenza viruses have the ability to agglutinate erythrocytes by binding to sialic acid receptors on host cells (C M Trombetta et al. 2018), resulting in decreased RBCs, HGB and HCT in influenza patients. Jarika (Jarika Makkoch et al. 2012)et al. observed the ability of antibodies against influenza A virus to bind to red blood cells through a hemagglutination inhibition test, and the results showed that a certain number of red blood cells in humans bound to antibodies against influenza A virus, which may explain why the RBC and HCT of influenza patients were lower than those of COVID-19 patients. However, a study reported (Marin-Mori et al. 2020) that COVID-19 patients also suffered from decreased coagulation function and anemia during hospitalization, which may result in a significantly lower number of RBC. Therefore, HCT is an important indicator for differentiating the two diseases.

To reduce the workload of clinicians and improve the rational utilization of medical resources, the establishment of an effective prediction model has important clinical significance. Currently, clinical decision models have been explored to validate the prognosis of SARS-CoV-2. Sun et al. established a prediction model including laboratory blood tests, clinical symptoms and radiology(Sun et al. 2020), Fabrizio et al. established a prediction model for diagnosing disease severity(Fabrizio Foieni et al. 2020), and Ma et al. established a prediction model for patient mortality based on laboratory findings(Ma X et al. 2020). The model established in this paper has the following two advantages. First, the model is concise and easy to understand and use. Second, this model was mainly used to distinguish COVID-19 from influenza A based on the typical laboratory findings selected comprehensively, including hematological and biochemical parameters, which indirectly provides a good clinical basis for diagnosis.

In summary, this study established a laboratory diagnostic model for COVID-19 and influenza A patients and identified more representative indicators for the segmentation of the two diseases. This model may provide better diagnostic clues and treatment plans for clinical practice that has certain clinical practicality.

**Conclusion**

Currently, there is a lack of effective drugs and vaccines to prevent and treat COVID-19 in the clinic. The winter time accelerates the spread of SARS-CoV-2 and influenza A virus. Due to the two diseases are extremly similar in clinical symptoms and transmission routes Therefore, it was necessary to effectively diagnose and treat these two diseases, prevent the co-infection of both viruses and identify COVID-19 and influenza A. In this study, A/G, TBIL and HCT were selected as highly specific indicators for differentiating these two diseases based on the diagnostic model established by laboratory findings of the COVID-19 and influenza A patients. External validation (AUC=0.844) proved the good applicability of the diagnostic model. Therefore, the above indicators can be used for the clinical diagnosis of COVID-19 and influenza A.

**Acknowledgments**

This work was supported by grants from Jilin Science and Technology Development Program (no. 20170623092TC-09, to Dr Jiancheng Xu; no. 20190304110YY to Dr Jiancheng Xu; no. 20200404171YY to Dr Qi Zhou) and the First Hospital Translational Funding for Scientific &Technological Achievements (no. JDYYZH-1902002 to Dr. Jiancheng Xu).

**Authors' contributions**

JX, ST, and DX conceived and designed the study. ST and DX collected and analyzed the data. YC, XS, JL and SL provided raw data and external validation data. ST and DX wrote the first draft of the manuscript. ST and JX revised the manuscript. All authors reviewed and edited the manuscript and approved the final version of the manuscript.

**Compliance with Ethical Standards**

**Conflicts of Interest:** The authors declare no conflicts of interest.

**Animal and Human Rights Statement**: This study was approved by the Ethics Committee of the First Hospital of Jilin University, Changchun Infectious Disease Hospital, Siping

Infectious Disease Hospital and Jilin Infectious Disease Hospital. Data were collected from the electronic patient record.

## Tables and Figures

**Table 1 The baseline characteristics of patients with COVID-19 and influenza A**

|  | COVID-19 (*n*=56) | Influenza A (*n*=54) | *P* |
|---|---|---|---|
| **Hospitalized time, d** | 18 (15-20) | 7 (5-10) | <0.001 |
| **Male** (%) | 31 (55.4) | 36 (66.7) | 0.246 |
| **Age, y** | 40 (28-51) | 63 (48-71) | <0.001 |
| **Clinical characteristics** | | | |
| **Fever** | 46 (82.1) | 29 (53.7) | 0.002 |
| **Cough** | 44 (78.6) | 19 (35.2) | <0.001 |
| **Hemoptysis** | 1 (1.8) | 1 (1.9) | 0.743 |
| **Headache** | 6 (10.7) | 2 (3.7) | 0.271 |
| **Chest pain** | 2 (3.6) | 2 (3.7) | 0.677 |
| **Fatigue** | 20 (35.7) | 9 (16.7) | 0.030 |
| **Muscle pain** | 9 (16.1) | 4 (7.4) | 0.238 |
| **Dyspnea** | 6 (10.7) | 11 (20.4) | 0.193 |
| **Abdominal pain or** | 6 (10.7) | 6 (11.1) | 0.593 |

Data are presented as median (*P*25-*P*75) or No. (%), y: year; d: day.

**Table 2 Laboratory Findings at Onset to Hospital Admission**

| Analytes | Reference interval | COVID-19 | Influenza A | P |
|---|---|---|---|---|
| | | Mean (P25-P75) | | |
| **Hematological Parameter** | | | | |
| White blood cells | 3.50-9.50 | 5.22 (3.90-6.30) | 7.33(5.44-9.40) | <0.001 |
| Neutrophils ($10^9$/L) | 1.80-6.30 | 3.45 (2.48-4.38) | 5.43 (3.73-6.59) | <0.001 |
| Lymphocyte ($10^9$/L) | 1.10-3.20 | 1.35 (0.90-1.66) | 1.20 (0.70-1.49) | 0.049 |
| Eosinophils ($\times 10^9$/L) | 0.02-0.52 | 0.03 (0.00-0.04) | 0.08 (0.00-0.12) | 0.094 |
| Basophil ($\times 10^9$/L) | 0.00-0.06 | 0.02 (0.00-0.03) | 0.03 (0.01-0.04) | 0.199 |
| Monocyte ($\times 10^9$/L) | 0.10-0.60 | 0.36 (0.24-0.50) | 0.59 (0.40-0.76) | <0.001 |
| Red blood cell | 4.30-5.80 | 4.69 (4.29-5.09) | 4.08 (3.40-4.66) | <0.001 |
| Hemoglobin, g/L | 130.00-175.00 | 145.00 (133.00-158.00) | 125.98 (104.00-142.50) | <0.001 |
| Hematocrit, L/L | 0.40-0.50 | 0.42 (0.39-0.45) | 0.37 (0.31-0.42) | <0.001 |
| MCV, fL | 82.0-100.0 | 89.68 (86.58-92.60) | 91.34 (87.15-93.55) | 0.201 |
| MCH, pg | 27.0-34.0 | 30.82 (30.00-32.03) | 30.90 (29.60-32.45) | 0.810 |
| MCHC, g/L | 316-354 | 344.69 (340.75-349.00) | 337.79 (327.50-347.50) | <0.001 |
| RDW, % | 11.00-16.00 | 11.65 (11.20-11.90) | 13.76 (12.55-14.65) | <0.001 |
| Platelets ($\times 10^9$/L) | 125.00-350.00 | 201.83 (159.25-230.50) | 193.08 (125.00-281.00) | 0.342 |
| MPV, fL | 6.5-12.0 | 9.40 (8.58-10.10) | 10.79 (9.70-11.55) | <0.001 |
| PDW, % | 9.0-17.0 | 12.81 (10.65-14.35) | 12.32 (10.50-12.90) | 0.216 |
| PCT, % | 0.108-0.282 | 0.19 (0.14-0.20) | 0.20 (0.14-0.29) | 0.485 |
| **Biochemical parameters** | | | | |
| AST, U/L | 13.00-40.00 | 29.30 (20.00-30.00) | 52.50 (18.20-78.35) | 0.041 |
| ALT, U/L | 7.00-50.00 | 34.23 (19.00-45.00) | 43.25 (12.80-63.95) | 0.617 |
| GGT, U/L | 7.00-60.00 | 34.30 (15.00-40.00) | 72.33 (18.50-113.70) | <0.001 |
| Cholinesterase, U/L | 5000-12000 | 8032 (6885-9045) | 4865 (3811-5828) | <0.001 |
| Total protein, g/L | 65.0-85.0 | 68.43 (65.00-72.00) | 60.60 (56.30-64.60) | <0.001 |
| Albumin, g/L | 40.0-55.0 | 44.30 (42.20-45.90) | 30.95 (26.55-35.70) | <0.001 |
| Globulin, g/L | 20.0-40.0 | 24.08 (21.60-26.70) | 29.65 (26.35-33.30) | <0.001 |
| A/G | (1.2-2.4)/1 | 1.89 (1.66-2.06) | 1.08 (0.86-1.18) | <0.001 |
| Total bilirubin, μmol/L | ≤23.00 | 10.00 (6.70-13.20) | 19.61 (9.80-25.50) | <0.001 |
| Direct bilirubin, μmol/L | ≤8.00 | 4.10 (3.10-5.10) | 8.78 (3.00-9.55) | <0.001 |
| Indirect bilirubin, | 5.10-21.40 | 5.87 (3.40-7.50) | 10.83 (6.45-13.15) | <0.001 |
| Urea nitrogen, mmol/L | 3.10-8.00 | 3.96 (3.07-4.75) | 6.12 (3.88-8.24) | <0.001 |
| Creatinine, μmol/L | 57.00-97.00 | 67.90 (57.50-77.00) | 70.08 (47.25-84.38) | 0.207 |
| $CO_2$-CP, mmol/L | 22.00-29.00 | 20.74 (18.9-22.4) | 25.78 (24.23-27.80) | <0.001 |
| Glucose, mol/L | 3.89-6.11 | 6.62 (5.53-6.79) | 6.38 (5.01-7.26) | 0.392 |
| Potassium,mmol/L | 3.50-5.30 | 4.14 (3.90-4.40) | 3.95 (3.60-4.30) | 0.036 |
| Sodium, mmol/L | 137.00-147.00 | 138.50 (137.00-140.00) | 136.74 (134.00-140.00) | 0.021 |
| Chloride, mmol/L | 99.00-110.00 | 99.73 (97.25-102.00) | 101.54 (98.00-104.00) | 0.030 |

Data are presented as means (P25-P75). MCV, Mean corpuscular volume; MCH, Mean corpuscular hemoglobin; MCHC, Mean corpuscular hemoglobin concentration; RDW, Red blood cell distribution width ;MPV, Mean platelet volume; MPV, Mean platelet volume; PDW, Platelet distribution width; PCT, Thrombocytocrit; AST, Aspartate aminotransferase; ALT, Alanine aminotransferase; GGT, γ-glutamyltranspeptidase; A/G, Albumin to Globulin ratio; $CO_2$-CP, Carbondioxide combining power.

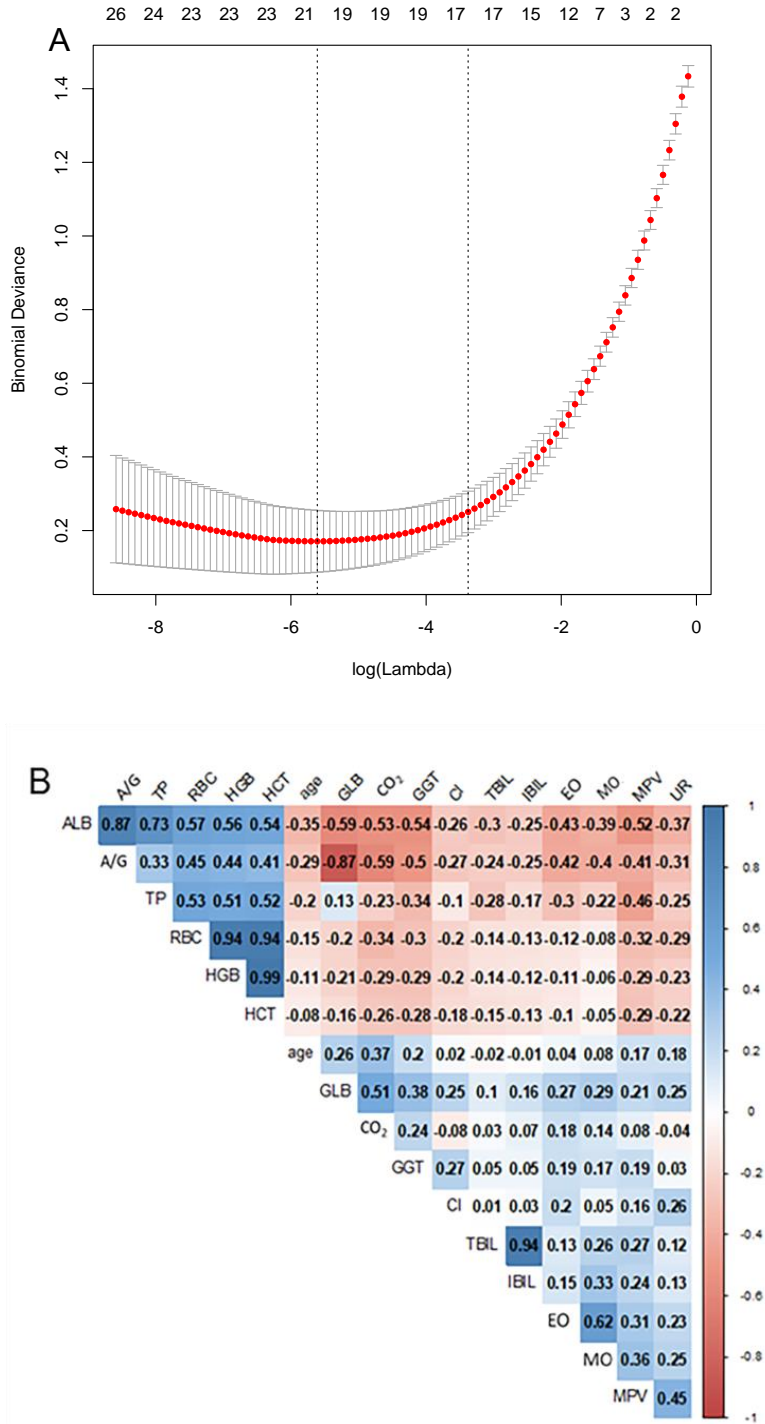

Fig.1 (A) Determination of the optimal value for tuning parameter λ in an elastic net model
(B) Correlation plot illustrating how the selected indices are correlated.

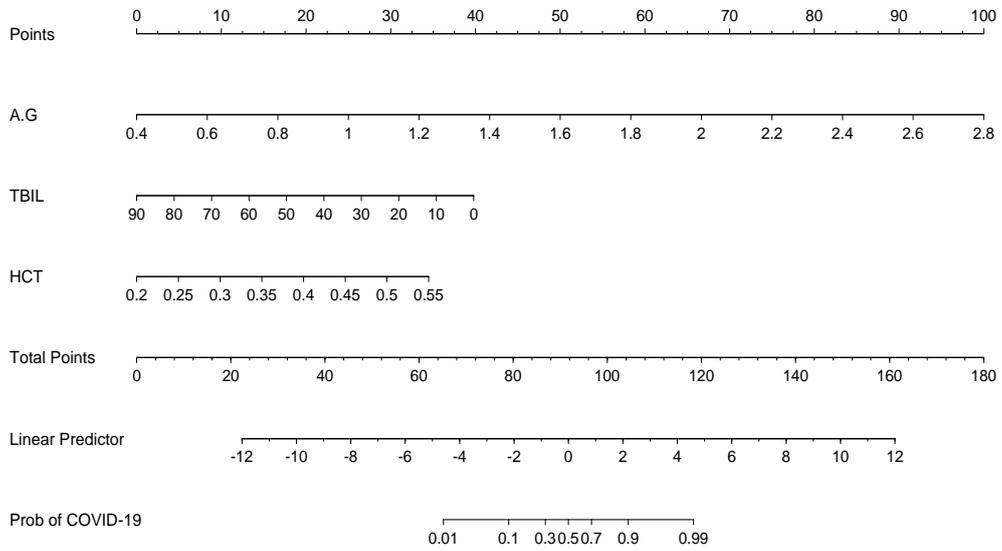

Fig.2　Nomogram showing the final model to discriminate COVID-19 and influenza A.

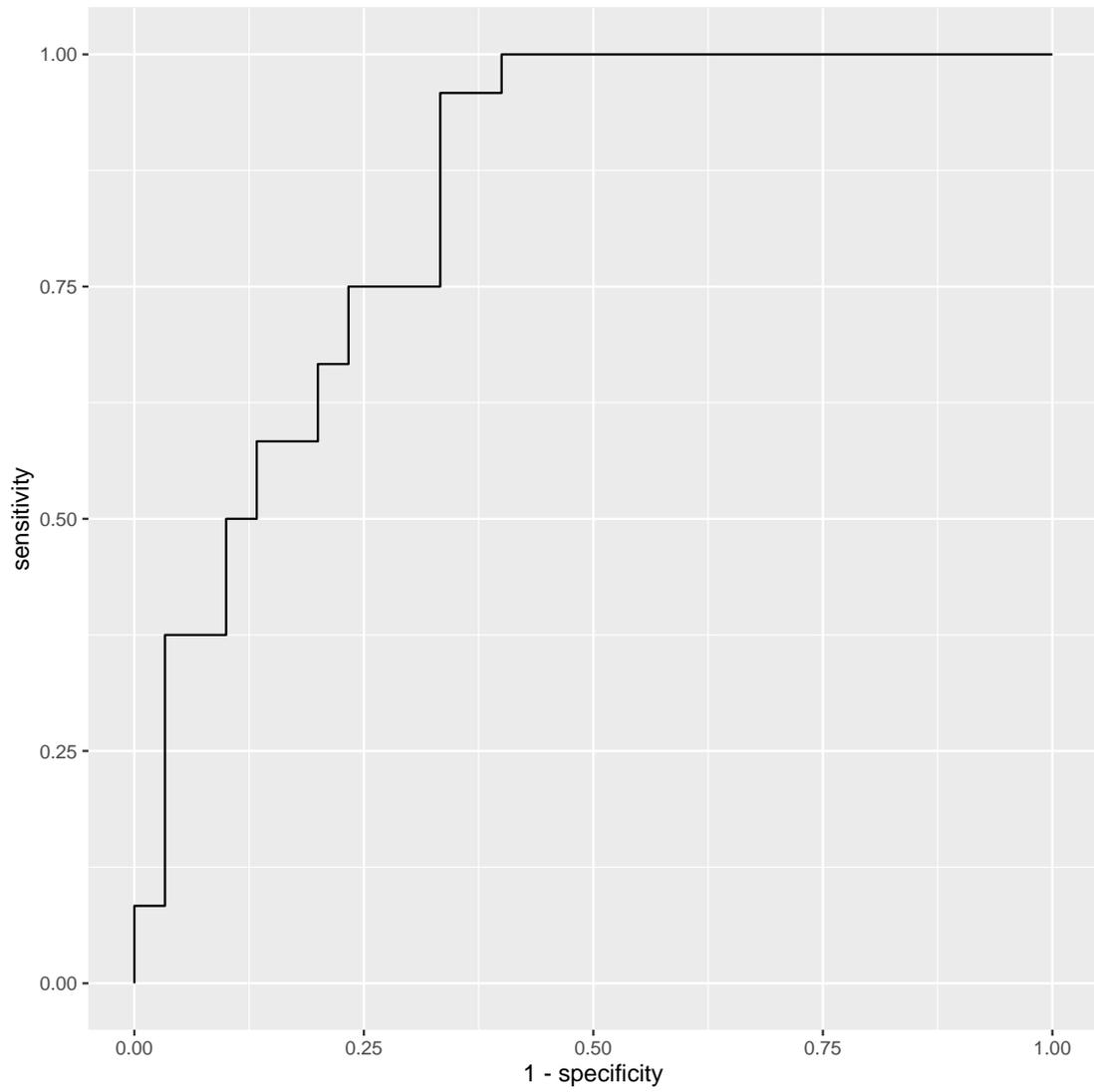

Fig.3 ROC curve of the final model on the external validation dataset.